\documentclass[aps,prb,twocolumn,groupedaddress,showpacs]{revtex4}

\usepackage{latexsym,amssymb,amsmath,amsfonts}
\usepackage{graphicx,color,pstricks}
\usepackage{epsfig,epsf,bm}

\definecolor{gray}{rgb}{0.7,0.7,0.7}

\begin{document}

\title{Tunneling between a topological superconductor and a Luttinger liquid}

\author{Yu-Wen Lee}
\email{ywlee@thu.edu.tw} \affiliation{Department of Physics, Tunghai University, Taichung, Taiwan, Republic of China}

\author{Yu-Li Lee}
\email{yllee@cc.ncue.edu.tw} \affiliation{Department of Physics, National Changhua University of Education, Changhua,
Taiwan, Republic of China}

\date{\today}

\begin{abstract}
 We study the quantum point contact between the topological superconductor and the helical Luttinger liquid. The effects
 of the electron-electron interactions in the helical Luttinger liquid on the low-energy physics of this system are
 analyzed by the renormalization group. Among the various couplings at the point contact which arise from the tunneling
 via the Majorana edge channel, the induced backscattering in the helical Luttinger liquid is the most relevant for
 repulsive interactions. Hence, at low temperatures, the helical Luttinger liquid is effectively cut into two separated
 half wires. As a result, the low-temperature physics is described by a fixed point consisting of two leads coupled to
 the topological superconductor, and the electrical transport properties through the point contact at low temperature and
 low bias are dominated by the tunneling via the Majorana edge channel. We compute the temperature dependence of the
 zero-bias tunneling conductance and study the full counting statistics for the tunneling current at zero temperature.
\end{abstract}


\maketitle

\section{Introduction}

One of the central issues in the study of condensed matter physics in recent years is to verify the existence of Majorana
fermions in various condensed matter systems. This tide of research has been partly triggered by the suggestion that the 
Majorana modes have great potential in the applications of fault-tolerant quantum computations.\cite{Kitave,Nayak} 
Theoretically, it was found that the Majorana edge states are supported by certain quantum Hall states as well as the 
$p_x+ip_y$-wave superconductors and superfluids.\cite{Moore-Read,Read-Green,Gurarie,MStone} More recently, Fu and 
Kane\cite{Fu-Kane-1} proposed that Majorana fermions can be created in the vortices of $s$-wave superconductors deposited 
on the surface of a three-dimensional topological insulator.\cite{Fu-Kane-2,Fu-Kane-3,Moore-Balents,HZhang,DHsieh,YXia} 
In particular, chiral Majorana fermion edge states can be created at the interface between a superconductor and the area 
gapped by ferromagnetic materials.\cite{Fu-Kane-1} Motivated by these theoretical proposals, the transport properties 
between the Majorana edge states and the (noninteracting) normal metallic lead were studied in a pioneering work,\cite{KTLaw} 
and it was found that the Majorana fermions induce resonant Andreev reflection from the lead to the grounded superconductor. 
The resulting linear tunneling conductance is $0$ or $2e^2/h$, depending on whether there is an even or odd number of 
vortices in the superconductor.

Another interesting proposal for the realization of Majorana modes was made in two seminal works in Refs.
\onlinecite{Lutchyn-Sau-DasSarma} and \onlinecite{Oreg-Refael-Open}, motivated by a model proposed by Kitaev.\cite{kitaev2} It
was found that the localized Majorana modes can exist at the end of a spin-orbit coupled nanowire subjected to a magnetic field
and proximate to an $s$-wave superconductor. Following these proposals, experimental evidence for such a Majorana edge mode
was obtained in indium antimonide quantum wires.\cite{Mourik} These developments initiated further theoretical studies on the
electrical transport properties of this system.\cite{flensberg,jdsau,golub,wimmer,stan,fisher,AG,Lutchyn-Sarabacz} In particular,
the effects of the electron-electron interactions in the metallic leads have been analyzed from a new perspective based on the
renormalization-group (RG) method.\cite{fisher,AG} The main conclusion is that among all possible local interactions in the
tunneling junction, the tunneling via the Majorana edge mode is the most relevant, and it drives the system all the way from
the perfect normal reflection fixed point to the perfect Andreev reflection fixed point. The latter controls the low-temperature
electrical transport properties, characterized by a universal zero-bias conductance at zero temperature ($2e^2/h$ for a single
lead with spinless fermions).

\begin{figure}
\begin{center}
 \includegraphics[width=0.8\columnwidth]{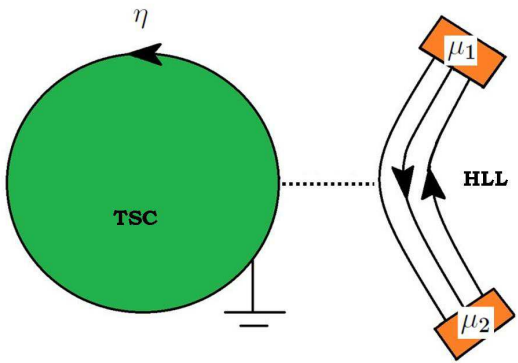}
 \caption{(Color online) A schematic picture of quantum point contact between the HLL and an island of TSC. At the edge of 
 the TSC, there is a branch of chiral Majorana fermions denoted by $\eta$. Electrons can tunnel between the HLL and the TSC 
 through the point contact. The superconductor is grounded. The chemical potentials at the two leads are chosen to be 
 identical, i.e., $\mu_1=\mu_2=V$.}
 \label{mlt1}
\end{center}
\end{figure}

\begin{figure}
\begin{center}
 \includegraphics[width=0.8\columnwidth]{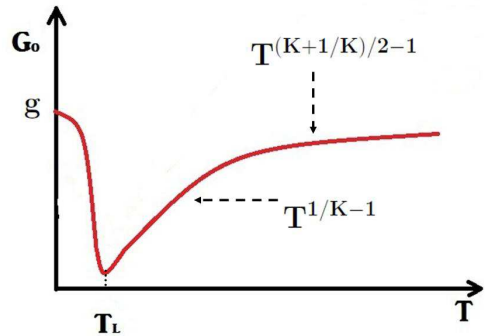}
 \caption{(Color online) A schematic temperature dependence of the tunneling conductance $G_0$ (in units of $e^2/h$) at 
 zero bias. For $T>T_L=|v_M|/L$, $G_0$ is a monotonously decreasing function with decreasing temperature $T$, where $L$ is 
 the length of the Majorana liquid. At high temperature, it behaves like $G_0\propto T^{\frac{1}{2}(K+1/K)-1}$, where $K$ 
 is the LL parameter. At low temperature, it is $G_0\propto T^{1/K-1}$ for $T>T_L$. For $T<T_L$, $G_0$ increases as 
 decreasing $T$ and reaches a universal value $g$ in units of $e^2/h$ at $T=0$ provided that there exists a Majorana 
 zero-energy state, where $g$ is a function of $K$.}
 \label{mlg1}
\end{center}
\end{figure}

In the present work, we study a quantum point contact (QPC) between a (chiral) Majorana liquid at the edge of a topological
superconductor (TSC) and a helical (spinless) Luttinger liquid (HLL),\cite{KM} as shown in Fig. \ref{mlt1}. Our work
intervenes between Refs. \onlinecite{KTLaw} and \onlinecite{fisher} in the sense that we investigate the interplay between
the electron-electron interaction effect in the metallic lead and the role of the propagating Majorana mode, instead of a
localized mode. The main results of our work are summarized in Fig. \ref{mlg1}. Our analysis focuses on the intermediate
temperature range below the superconducting gap such that the finite-size effects of both the HLL and the Majorana liquid can be
ignored. Under this condition, our RG analysis in the weak-tunneling limit shows that, unlike the case with a localized Majorana
end mode, the tunneling between the Majorana liquid and the HLL is inhibited at low energy and the system is driven to a new
fixed point (referred to as the fixed point $I$ in the main text) in which the HLL is effectively cut into two separated pieces
by a local backscattering potential at the QPC. (This local backscattering potential is produced by the proximity to the TSC, 
and is allowed because the time-reversal symmetry is broken by the $p_x+ip_y$-wave TSC.) Moreover, we also performed a scaling
analysis regarding the stability of the perfect Andreev reflection fixed point (referred to as the fixed point $A$) in the 
strong-tunneling limit and found that it is unstable toward the weak-tunneling regime. Thus, the possibility of the existence 
of a nontrivial quantum critical point in the intermediate-tunneling-strength regime, which separates fixed point $I$ from fixed 
point $A$, is unlikely. Because the low-temperature and high-temperature physics of this system are controlled by different
fixed points, we obtain a two-stage scaling of the tunneling conductance with temperature, dictated by different exponents as
shown in Fig. \ref{mlg1}. Since in real systems, the Majorana liquid is most likely to be of the mesoscopic length
scale,\cite{Fu-Kane-1} we expect that at temperatures below the scale $T_L$ set by the level spacing of the Majorana modes,
the Majorana liquid will behave like a localized mode and the system will eventually experience a crossover to the model studied
in Ref. \onlinecite{AG}, i.e. two metallic leads coupled to a single Majorana mode. Therefore, the tunneling via the Majorana
mode will become a relevant perturbation, and the tunneling conductance of the system at zero temperature will saturate at a
universal value depending on the LL parameter, as long as there is a Majorana edge state whose energy matches the Fermi level of
the lead.\cite{AG}

In the remaining sections, we present systematical theoretical analysis which supports the above conclusions and compare
our results with previous works in the final section.

\section{The Model}

We consider a QPC between the grounded TSC and the HLL as shown in Fig. \ref{mlt1}. This system can be described by the
Hamiltonian $H=H_{LL}+H_{\eta}+\delta H$, where
\begin{eqnarray}
 H_{LL} &=& \sum_{m=\pm} \! \int \! dx\left[\psi^{\dagger}_m(-imv_F\partial_x\psi_m)+g_1J_mJ_m\right] \nonumber \\
 & & +g_2\! \int \! dxJ_+J_-  \label{qpch1}
\end{eqnarray}
is the Hamiltonian of the HLL, and
\begin{equation}
 H_{\eta}=\frac{v_M}{2}\! \int^{L/2}_{-L/2} \! dx\eta(-i\partial_x\eta) \label{qpch11}
\end{equation}
is the Hamiltonian of the chiral Majorana liquid. In the above, $v_F$ and $|v_M|$ are the Fermi velocity in the HLL and the
speed of the Majorana fermion at the edge of the TSC, respectively. The operators $\psi_+$ and $\psi_-$, which describe
the right and leftmover in the HLL, obey the canonical anticommutation relations. $J_m=\psi_m^{\dagger}\psi_m$ with
$m=\pm$ are the current operators. Moreover, we take the normalization of the Majorana fermion $\eta$ so that it satisfies
the anticommutation relation
\begin{equation}
 \{\eta(x),\eta(y)\}=\delta(x-y) \ . \label{cml1}
\end{equation}
In the following, we shall focus on the limit $L\rightarrow +\infty$ and comment on the finite $L$ effects later.

$\delta H$ describes various couplings at the point contact. By keeping the most relevant terms and setting the QPC to be
located at $x=0$, $\delta H$ can be written as
\begin{eqnarray}
 \delta H &=& \! \sum_{m=\pm} \! \left\{-i\bar{t}\eta(0) \! \left[\xi_m\psi_m(0)+\mathrm{H.c.}\right]+2u_1\psi^{\dagger}_m
 \psi_m(0)\right\} \nonumber \\
 & & +\left[u_2\psi^{\dagger}_+\psi_-(0)+\Delta\psi_+\psi_-(0)+\mathrm{H.c.}\right] , \label{qpch12}
\end{eqnarray}
where $\bar{t},u_2>0$, $u_1$ is real, and $|\xi_m|=1$. The $\bar{t}$ term describes the tunneling between the TSC and the HLL
via the Majorana edge channel, the $u_1$ term describes the local chemical potential variation due to the presence of the
QPC, the $u_2$ term accounts for the backscattering in the HLL arising from the contact to the TSC (which is allowed in the
present case on account of the breaking of time-reversal symmetry by the TSC), and the $\Delta$ (Cooper-pairing) term is
induced by the proximity to the superconductor.

\section{The scaling analysis}

We now study the effects of $\delta H$ on the decoupled fixed point, which corresponds to $\delta H=0$. We first ignore the
electron-electron interactions in the HLL, i.e. setting $g_1=0=g_2$. Then, the scaling dimensions of the various terms in $
\delta H$ around the decoupled fixed point are $D[\bar{t}]=1=D[u_1]=D[u_2]=D[\Delta]$. That is, all are marginal operators.
This is different from the coupling of a lead to the TSC via a single Majorana edge mode. For the latter, the $\bar{t}$ term
is already a relevant perturbation for free electrons.\cite{fisher} This distinction arises from the nontrivial scaling
dimension ($D[\eta]=1/2$) acquired by the Majorana fermion $\eta$ in the limit $L\rightarrow +\infty$.

\subsection{The weak-tunneling regime}
\label{wtr}

To take into account the effects of the electron-electron interactions (the $g_1$ and $g_2$ terms), we employ the method of
bosonization. Using the bosonization formula\cite{GNT},
$\psi_{\pm}(x)=\frac{\gamma}{\sqrt{2\pi a_0}}\exp{[\pm i\sqrt{4\pi}\phi_{\pm}(x)]}$, where $a_0$ is a short-distance cutoff
and $\gamma$ is the Klein factor with the normalization $\gamma^2=1$, $H_{LL}$ can be written as
\begin{equation}
 H_{LL}=\frac{v}{2} \! \int \! dx \! \left[K(\partial_x\Theta)^2+\frac{1}{K}(\partial_x\Phi)^2\right] , \label{qpch13}
\end{equation}
where $\Phi=\phi_-+\phi_+$, $\Theta=\phi_--\phi_+$, and $K$ is the LL parameter. For the repulsive (attractive) 
interactions, $K<1$ ($K>1$). On the other hand, the bosonized form of $\delta H$ is given by
\begin{eqnarray*}
 \delta H &=& -\frac{i\bar{t}\eta (0)\gamma}{\sqrt{2\pi a_0}} \! \left[\xi_+e^{i\sqrt{4\pi}\phi_+(0)}+\xi_
 -e^{-i\sqrt{4\pi}\phi_-(0)}+\mathrm{H.c.}\right] \\
 & & +\frac{2u_1}{\sqrt{\pi}}\partial_x\Phi(0)-\frac{u_2}{\pi a_0}\sin{\! \left[\sqrt{4\pi}\Phi(0)\right]} \\
 & & + \frac{|\Delta|}{\pi a_0}\sin{\! \left[\sqrt{4\pi}\Theta(0)-\alpha\right]} ,
\end{eqnarray*}
where we have written $\Delta$ as $\Delta=|\Delta|e^{i\alpha}$. The $u_1$ term can be removed by the transformation
$\Phi(x)\rightarrow\Phi(x)-\frac{Ku_1}{\sqrt{\pi}v}\mbox{sgn}(x)$, and $\delta H$ becomes
\begin{eqnarray}
 \delta H \! \! &=& \! \! -\frac{i\bar{t}\eta (0)\gamma}{\sqrt{2\pi a_0}} \! \left[\xi_+e^{i\sqrt{4\pi}\phi_R(0)} \!
 +\xi_-e^{-i\sqrt{4\pi}\phi_L(0)} \! +\mathrm{H.c.}\right] \nonumber \\
 \! \! & & \! \! -\frac{u_2}{\pi a_0}\sin{\! \left[\sqrt{4\pi}\Phi(0)\right]} \! +\frac{|\Delta|}{\pi a_0}
 \sin{\! \left[\sqrt{4\pi}\Theta(0)-\alpha\right]} . ~~~~\label{qpch14}
\end{eqnarray}

The scaling dimensions of various terms in $\delta H$ around the decoupled fixed point are given by
$D[\bar{t}]=\frac{1}{4}(K+1/K)+1/2$, $D[u_2]=K$, and $D[\Delta]=1/K$. Hence, we reach the following conclusions: (i) for 
repulsive interactions ($K<1$), the $u_2$ term is relevant while the $\Delta$ term is irrelevant; (ii) for attractive 
interactions ($K>1$), the $\Delta$ term is relevant while the $u_2$ term is irrelevant. In both cases, the $\bar{t}$ 
term is irrelevant and the $u_1$ term is marginal. In this paper, we focus on the case with $K<1$.

For $K<1$, the $u_2$ term will flow to the strong-coupling regime, and we expect that the low-energy physics is determined
by the fixed-point Hamiltonian
\begin{equation}
 H_1=H_{LL}+H_{\eta}-\frac{u_2}{\pi a_0}\sin{\! \left[\sqrt{4\pi K}\tilde{\Phi}(0)\right]} \ , \label{qpch2}
\end{equation}
where $\tilde{\Phi}=\Phi/\sqrt{K}$. $H_1$ takes the form of the Hamiltonian of a spinless LL with an impurity backscattering
term at $x=0$. Since $K<1$, this term is always relevant and cuts the wire into two separate pieces at $x=0$.\cite{KF}
This leads to the boundary condition $\tilde{\Phi}(0)=C_1=\frac{\pi/2}{\sqrt{4\pi K}}$. However, we still have to examine
the stability of the fixed point described by $H_1$ with $u_2\rightarrow +\infty$, which will be denoted by $I$.

To examine the stability of the fixed point $I$, we consider the possible perturbations around it. They are the $\bar{t}$
term, which becomes $-i\bar{t}\eta(0)\gamma \! \left[\xi e^{-i\sqrt{\pi}\Theta(0)}+\mathrm{H.c.}\right]$ by taking into
account the boundary condition at the fixed point $I$, the $\Delta$ term, as well as the tunneling between the two separated
half wires (denoted by $A$ and $B$) $\lambda\hat{O}$, where $\hat{O}=\Psi_A^{\dagger}(0)\Psi_B(0)+\mathrm{H.c.}$ and
$\Psi_{A(B)}$ is the fermion field referred to region $A$ ($B$). The scaling dimensions of the operators involving the
$\Theta$ field can be calculated by the action for the half wire in the imaginary-time formulation
\begin{eqnarray*}
 S_1=\frac{K}{2v} \! \int^{\beta}_0 \! \! d\tau \! \int^{+\infty}_0 \! dx \! \left[(\partial_{\tau}\Theta)^2
 +v^2(\partial_x\Theta)^2\right] ,
\end{eqnarray*}
yielding $D[\bar{t}]=1/(2K)+1/2$, $D[\Delta]=2/K$, and $D[\lambda]=1/K$. Hence, all terms are irrelevant for $K<1$. That is,
the fixed point $I$ is stable for repulsive interactions. Moreover, the leading irrelevant operator (LIO) around this fixed
point is the $\bar{t}$ term.

To sum up, the low-energy physics of the QPC between the TSC and the HLL is controlled by the fixed point $I$, which is
composed of two half wires and the chiral Majorana liquid on the edge of the TSC. The tunneling between the two half wires
and the Majorana liquid gives rise to the leading temperature dependence of thermodynamics at low temperatures. We shall see
later that the same coupling also dominates the electrical transport properties of the point contact at low temperatures
(bias).

\subsection{The strong-tunneling regime}

The previous analysis holds only in the weak-tunneling regime. It is still possible that the low-energy physics in the large
$\bar{t}$ limit is described by the fixed point of perfect Andreev reflection (denoted by $A$), as in the case of coupling
to a single Majorana mode. We show that this is not the case, and the fixed point $A$ is unstable toward the weak-tunneling
regime.

We start with a tight-binding model described by the Hamiltonian $H=H_0+H_{int}+H_t$, where
\begin{eqnarray*}
 H_0 &=& -\frac{w_1}{2} \! \sum_{j=-\infty}^{+\infty} \! \! \left(c_{j+1}^{\dagger}c_j+\mathrm{H.c.}\right) \! -\frac{iw_2}
 {2} \! \sum_{j=-\infty}^{+\infty} \! \eta_j\eta_{j+1} \ , \\
 H_{int} &=& U \! \sum_{j=-\infty}^{+\infty} \! \! \left(c^{\dagger}_jc_j-\frac{1}{2}\right) \! \! \left(c^{\dagger}_{j+1}
 c_{j+1}-\frac{1}{2}\right) , \\
 H_t &=& \frac{it}{\sqrt{2}}\eta_0(\xi c_0+\mathrm{H.c.}) \ .
\end{eqnarray*}
In the above, the LL is described by the repulsive Hubbard model of spinless fermions, with $c_j$, $c_j^{\dagger}$
satisfying the canonical anticommutation relations and $U>0$. Without loss of generality, we may take $w_1,w_2,t>0$.

We consider the limit that $w_1/t,w_2/t,U/t\ll 1$. Setting $w_1=0=w_2=U$ in $H$ gives the ``strong-tunneling" Hamiltonian
$H_t$. To diagonalize $H_t$, we write $c_0$ as $\xi c_0=\frac{1}{\sqrt{2}}(\gamma+i\bar{\gamma})$, where $\gamma$ and
$\bar{\gamma}$ are Majorana fermions with the normalization $\gamma^2=1/2=\bar{\gamma}^2$. Then, $H_t$ can be written as
\begin{eqnarray*}
 H_t=it\eta_0\gamma \ .
\end{eqnarray*}
We may define the fermion operator $\psi\equiv\frac{1}{\sqrt{2}}(\eta_0+i\gamma)$ so that $H_t$ becomes
\begin{eqnarray*}
 H_t=t\psi^{\dagger}\psi \ ,
\end{eqnarray*}
up to a constant term. It is clear that the ground state of $H_t$ is the zero $\psi$-fermion state, while the excited state
of $H_t$ consists of one $\psi$ fermion. The two states are separated by an energy gap $E_g=t$.

Now we consider the perturbations for $H_t$ by turning on $w_1$ and $w_2$. (For simplicity, we consider only the
noninteracting case $U=0$.) They are given by $V+\bar{V}$, where
\begin{eqnarray*}
 V \! \! &=& \! \! -\frac{w_1}{2\sqrt{2}}[\xi\gamma(c_1+c_{-1})+\mathrm{H.c.}]-\frac{iw_2}{2}\eta_0(\eta_1-\eta_{-1}) \ ,
 \\
 \bar{V} \! \! &=& \! \! -\frac{w_1}{2\sqrt{2}}[-i\xi\bar{\gamma}(c_1+c_{-1})+\mathrm{H.c.}] \ .
\end{eqnarray*}
Because $[H_t,\bar{V}]=0$, the inclusion of $\bar{V}$ will not disrupt the entanglement between the $\eta_0$ and $\gamma$
Majorana modes in the ground state of $H_t$. Therefore, if $V=0$, we expect that this model will renormalize to the fixed
point $A$ at low energies.

The leading perturbation $\delta H$ to the fixed point $A$ is produced by $V$. With the help of the Schrieffer-Wolff
transformation,\cite{SW} we find that
\begin{eqnarray*}
 \delta H=\frac{w_1w_2}{8\sqrt{2}t}[\xi (c_1+c_{-1})-\mathrm{H.c.}](\eta_1-\eta_{-1})+\mathrm{H.c.} \ .
\end{eqnarray*}
Using the boundary condition for the fixed point $A$, i.e. $\Theta(x=0)=0$ mod $\sqrt{\pi}$, this expression suggests the
following bosonized form of $\delta H$:
\begin{eqnarray*}
 \delta H\sim -i\eta (x=0)\gamma^{\prime}\sin{[\sqrt{\pi}\Phi(0)]} \ ,
\end{eqnarray*}
where $\gamma^{\prime}$ is the Klein factor. Taking into account the $U$ term, the scaling dimension of the operator
$\sin{[\sqrt{\pi}\Phi(0)]}$ can be calculated in terms of the action for the half wire in the imaginary-time formulation
\begin{eqnarray*}
 S_2=\frac{1}{2vK} \! \int^{\beta}_0 \! \! d\tau \! \int^{+\infty}_0 \! dx \! \left[(\partial_{\tau}\Phi)^2
 +v^2(\partial_x\Phi)^2\right] ,
\end{eqnarray*}
leading to $D[\delta H]=(K+1)/2$. Since $D[\delta H]<1$ for $K<1$, it is a relevant perturbation such that the fixed point
$A$ is unstable.

\section{Electrical transport}

Based on the above analysis, we may compute the electrical transport properties for this QPC by applying a bias between
the HLL and the edge of the TSC. We will consider the weak-tunneling limit so that a perturbative expansion in $\bar{t}$
is valid.

\subsection{Tunneling conductance}

Applying a bias $V$ between the HLL and the edge of the TSC will generate a tunneling current $I_t$ through the point contact.
The quantity we would like to compute is the tunneling conductance $G=dI_t/dV$. This can be done as follows.

The action of the system in the real-time formulation can be written as
\begin{eqnarray*}
 S=\! \int \! \! dt \! \left[\sum_{m=\pm} \! \int \! \! dx\psi_m^{\dagger}(i\partial_t+\mu)\psi_m-H\right] ,
\end{eqnarray*}
where $\mu=-eV$. We perform a time-dependent gauge transformation $\psi_m\rightarrow e^{i\mu t}\psi_m$, leading to
$\Phi\rightarrow\Phi$ and $\Theta\rightarrow\Theta+\frac{eVt}{\sqrt{\pi}}$. Consequently, the dependence on $V$ will appear
only in the $\bar{t}$ and the $\Delta$ terms. The tunneling current operator is given by
$\hat{I}=-\frac{\delta H}{\delta a}=\hat{I}_1+\hat{I}_2$, where $a=-Vt$ and
\begin{eqnarray}
 \hat{I}_1 &\propto& i\gamma\eta (0)e^{-ieVt} \! \left[\xi_+e^{i\sqrt{4\pi}\phi_+(0)} \! +\xi_-e^{-i\sqrt{4\pi}\phi_-(0)}
 \right] \nonumber \\
 & & +\mathrm{H.c.} \ , \nonumber \\
 \hat{I}_2 &\propto& \cos{\! \left[\sqrt{4\pi}\Theta(0)-\alpha+eVt\right]} . \label{qpci1}
\end{eqnarray}
The tunneling current is then given by $I_t=\langle\hat{I}\rangle$.

The scaling dimensions $D_1$ and $D_2$ for the operator $\hat{I}_1$ and $\hat{I}_2$ are given by $D_1=D[\bar{t}]$ and
$D_2=D[\Delta]$, respectively. Since $D_1<D_2$ for $K<1$, we have $G(T=0)\propto V^{2D_1-2}$ at low bias, and
$G_0\equiv G(V=0)\propto T^{2D_1-2}$ at low temperatures. Therefore, the temperature dependence of $G_0$ reveals the RG
flow of the $\bar{t}$ term. Since the high-energy and low-energy physics of the system are controlled by the decoupled
fixed point and the fixed point $I$, we conclude that $G_0\propto T^{\frac{1}{2}(K+1/K)-1}$ at high temperature and
$G_0\propto T^{1/K-1}$ at low temperature. Moreover, $G_0$ is a monotonously decreasing function with decreasing
temperature.

In practice, the Majorana liquid has a finite length $L$, which introduces a new characteristic energy scale $T_L=|v_M|/L$.
The above results hold only when $T_L\ll T\ll\mbox{min}\{E_F,\Delta_g\}$, where $E_F$ is the Fermi energy of the HLL and
$\Delta_g$ is the gap of the TSC. For $T<T_L$, this problem becomes one with two leads coupled to a single Majorana
mode. It turns out that the physics at low temperatures is controlled by a nontrivial critical point,\cite{AG} which gives
rise to a universal conductance $G_0=g(K)e^2/h$ at $T=0$, where $g$ is a universal function of $K$.\cite{foot1} A schematic
temperature dependence of $G_0$ is shown in Fig. \ref{mlg1}.

\subsection{Full counting statistics}

It has been known that a measurement of current noise reveals information on the studied system not present in the average
current.\cite{schottky} Further information can be obtained by studying the higher moments (cumulants) of charge transfer,
or in general, the full counting statistics (FCS).

To extract the cumulants of charge transfer through the QPC, we may calculate the cumulant generating function
$W(\chi)=\ln{[Z(\chi)]}$. In the present case, $Z(\chi)$ is defined as
\begin{equation}
 Z(\chi)=\! \left\langle\exp{\! \left[\oint \! dt\chi(t) \! \int^{+\infty}_{-\infty} \! dx \! \sum_{m=\pm}\psi_m^{\dagger}\psi_m\right]}
 \! \right\rangle , \label{qpcz1}
\end{equation}
where the time integral is taken over the Keldysh contour and $\chi(t)$ is the counting field. In order that $Z(\chi)\neq 1$,
we must choose $\chi_+(t)\neq\chi_-(t)$, where the subscripts $+$ and $-$ refer to the forward and the backward branches of
the closed time contour, respectively. Here, we choose $\chi_+(t)=\chi\delta(t-\mathcal{T})$ and $\chi_-(t)=0$, where
$\mathcal{T}$ is the measurement time. For large $\mathcal{T}$, $W$ is proportional to $\mathcal{T}$ so that
\begin{eqnarray}
 I_t &=& \! \left.ie\lim_{\mathcal{T}\rightarrow +\infty}\frac{1}{\mathcal{T}}\frac{\delta W}{\delta\chi}\right|_{\chi=0} ,
 \label{qpci2} \\
 S(0) &=& \! \left.-e^2\lim_{\mathcal{T}\rightarrow +\infty}\frac{1}{\mathcal{T}}\frac{\delta^2W}{\delta\chi^2}
 \right|_{\chi=0} , \label{qpci21}
\end{eqnarray}
where $I_t$ is the tunneling current and $S(0)$ is the noise power of $I_t$ at zero frequency\cite{foot2}.

To proceed, we perform a time-dependent gauge transformation $\psi_m\rightarrow e^{i\theta (t)}\psi_m$, where
$\theta_+(t)=-\chi\Theta(\mathcal{T}-t)$ and $\theta_-(t)=0$. Then, $Z(\chi)$ can be written as
\begin{equation}
 Z(\chi)=\! \int \! \! D[\eta]D[\gamma]D[\tilde{\Theta}]\exp{\! \left[\oint \! dt (L_0+L_t)\right]} , \label{qpcz11}
\end{equation}
where $\tilde{\Theta}=\sqrt{K}\Theta$,
\begin{eqnarray}
 L_0 &=& \! \sum_{l=1,2}\frac{1}{2v} \! \int^{+\infty}_0 \! dx \! \left[(\partial_t\tilde{\Theta}_l)^2-v^2
 (\partial_x\tilde{\Theta}_l)^2\right] \nonumber \\
 & & +\frac{i}{2} \! \int^{+\infty}_{-\infty} \! dx\eta(\partial_t+v_M\partial_x)\eta +\frac{i}{4}\gamma\partial_t\gamma
 \ ,~~~~\label{qpcs1}
\end{eqnarray}
describes the fixed point $I$, and
\begin{equation}
 L_t=-i\bar{t}\eta (t,0)\gamma(t) \! \sum_{l=1,2}\cos{\! \left[\sqrt{\frac{\pi}{K}}\tilde{\Theta}_l(t,0)+\alpha(t)\right]}
 , \label{qpcs11}
\end{equation}
gives the LIO. In Eq. (\ref{qpcs11}), $\alpha (t)=eVt-\theta(t)$.

At low temperatures (bias), we may calculate $W$ in terms of the perturbative expansion in $\bar{t}$. To $O(\bar{t}^2)$,
we find that
\begin{equation}
 W=\frac{a_0^{1/K}v^{1-1/K}K}{2|v_M|\Gamma(1/K)}\mathcal{T}\bar{t}^2|\omega_0|^{1/K} \! \! \left[
 e^{-i\mbox{sgn}(\omega_0)\chi}-1\right] \label{qpci22}
\end{equation}
at $T=0$, where $\omega_0=eV$ and $a_0$ is the short-distance cutoff. Inserting Eq. (\ref{qpci22}) into Eqs. (\ref{qpci2})
and (\ref{qpci21}), we get the tunneling current at $T=0$:
\begin{equation}
 I_t=\frac{ea_0^{1/K}v^{1-1/K}K}{2|v_M|\Gamma(1/K)}\bar{t}^2\mbox{sgn}(\omega_0)|\omega_0|^{1/K} ,
 \label{qpci23}
\end{equation}
and the noise power of the tunneling current at zero frequency:
\begin{equation}
 S(0)=\frac{e^2a_0^{1/K}v^{1-1/K}K}{2|v_M|\Gamma(1/K)}\bar{t}^2|\omega_0|^{1/K} . \label{qpci24}
\end{equation}

A few comments regarding the above results are in order. First of all, the exponents of $|\omega_0|$ in Eqs. (\ref{qpci23}) 
and (\ref{qpci24}) are determined by the scaling dimension of $L_t$ around the fixed point $I$. For the usual
metal-superconductor junction, the current at $T=0$ due to the single-electron tunneling is suppressed by the superconducting 
(SC) gap $\Delta_g$ such that $I_t=0$ for $|\omega_0|<2\Delta_g$. In the present case, the low-energy states are allowed at 
the edge of the TSC, which turns the suppression of the tunneling current into a power-law behavior. Next, the Fano factor 
for the tunneling current is of the value
\begin{equation}
 F\equiv\frac{S(0)}{e|I_t|}=1 \ , \label{qpci25}
\end{equation}
instead of $2$ for the local Andreev reflection. It implies the single-electron tunneling through the QPC at low bias. This
is possible in the present case because of the overlapping between the wavefunction of electrons in the wire and that of
Majorana fermions in the TSC.

\section{Conclusion}

To sum up, we study a QPC between the chiral Majorana edge states of a TSC and the HLL. Our RG analysis predicts a
nonmonotonic behavior of the zero-bias tunneling conductance with temperature and the power-law temperature
dependence in the intermediate temperature range. Since tunneling into the usual nontopological superconductors (or
insulators) will make the tunneling conductance decrease exponentially with decreasing temperature due to the bulk gap in
the energy spectrum, the behavior of the tunneling conductance with temperature we found is an indication of the existence
of gapless states at the edge of the TSC. Moreover, we evaluate perturbatively the FCS of the tunneling current at zero
temperature, and use the corresponding result to extract the tunneling current and its noise power at zero frequency.

Previous studies on a distinct TSC-HLL tunneling junction shows that the low-temperature transport properties are controlled
by a perfect Andreev reflection fixed point\cite{fisher,Lutchyn-Sarabacz}, while in our case, it is the fixed point $I$
which controls the low-energy physics. Experimentally, the distinction between these two fixed points can also be revealed
by the corresponding Fano factors, which reflect the nature of the transported charges. Specifically, the Fano factor
obtained in our case is $1$ rather than $2$, as expected for a tunneling process dominated by Andreev reflections.\cite{KTLaw}
The origin of this difference is twofold: First of all, the scaling dimension of the tunneling via the Majorana edge states
is increased in our case by the propagating nature of Majorana fermions. Next, the metallic lead in the system studied in
Refs. \onlinecite{fisher} and \onlinecite{Lutchyn-Sarabacz} is essentially a half wire. Therefore, upon folding, the local
potential scattering in the tunneling junction contains only the forward scattering of electrons, whose sole effect is to
change the phase of the incident electron and does not affect the low-energy physics of the system qualitatively. On the
other hand, the potential scattering term in our system also contains the backscattering of electrons, which effectively cuts
the wire into two semi-infinite pieces at low energies.\cite{KF} We would like to stress that such a term must be included
not only because the time-reversal symmetry is broken by the $p_x+ip_y$-wave TSC, but also because it will inevitably be
generated at the energy scale below the SC gap through the virtual tunneling between the electrons in the HLL and the gapped
quasiparticles in the TSC. Another similar tunneling junction between two metallic leads and the end of a TSC wire via the
localized Majorana mode was analyzed in Ref. \onlinecite{AG}. In that case, a local backscattering can also be generated by
renormalization. However, the tunneling via the Majorana end mode is still the most relevant perturbation, which leads to
different low-energy physics from those presented here.

Our analysis in the present work can be extended readily to study the electron-electron interaction effect on the system
considered in Ref. \onlinecite{KTLaw}, which is a tunneling junction between a chiral Majorana liquid of finite size and
the metallic lead. According to the scaling analysis presented in Sec. \ref{wtr}, it is easy to see that the zero-bias
tunneling conductance in that system will also exhibit a nonmonotonic behavior with decreasing temperature. As the
temperature is lowered, the zero-bias conductance will decrease toward zero in the form of $T^{1/K-1}$, as shown by the 
second stage behavior presented in Fig. \ref{mlg1}. When $T\ll T_L$, the tunneling conductance will increase as the 
temperature decreases further and will eventually be saturated at the value of $2e^2/h$, as long as there is a zero-energy 
Majorana mode in the TSC, as predicted in Ref. \onlinecite{fisher}. This nonmonotonic scaling behavior of the tunneling 
conductance with temperature is in sharp contrast to that for tunneling into a single Majorana end mode.\cite{fisher,Lutchyn-Sarabacz} 
In the latter case, the tunneling conductance is expected to rise immediately as the temperature is lowered. The subtle 
difference between these two cases should serve as a guide in future experiments in identifying the possible Majorana edge 
states of TSC. Moreover, a complete analysis regarding the interplay between the effects of finite bias (temperature) and 
the finite size of the Majorana liquid is a topic for future study.

\acknowledgments

The work of Y.-W.L. is supported by the National Science Council of Taiwan under Grant No. NSC 102-2112-M-029 -002 -MY3.


\end{document}